# Isogeometric finite element analysis of laminated composite plates based on a four variable refined plate theory.

**Loc V. Tran**[1], **Chien H. Thai**[1], **Buntara S. Gan**[2] **and H. Nguyen-Xuan**[1,3*]

[1] Division of Computational Mechanics, Ton Duc Thang University Ho Chi Minh City, Vietnam

[2] Department of Architecture, College of Engineering, Nihon University, Koriyama City, Fukushima Prefecture, Japan

[3] Department of Mechanics, Faculty of Mathematics & Computer Science, University of Science Ho Chi Minh City, Vietnam

**Abstract**

In this paper, a novel and effective formulation based on isogeometric approach (IGA) and Refined Plate Theory (RPT) is proposed to study the behavior of laminated composite plates. Using many kinds of higher-order distributed functions, RPT model naturally satisfies the traction-free boundary conditions at plate surfaces and describes the non-linear distribution of shear stresses without requiring shear correction factor (SCF). IGA utilizes the basis functions, namely B-splines or non-uniform rational B-splines (NURBS), which achieve easily the smoothness of any arbitrary order. It hence satisfies the $C^1$ requirement of the RPT model. The static, dynamic and buckling analysis of rectangular plates is investigated for different boundary conditions. Numerical results show high effectiveness of the present formulation.

## 1. Introduction

Laminated composite plates are being increasingly used in various fields of engineering such as aircrafts, aerospace, vehicles, submarine, ships, buildings, etc, because they possess many favorable mechanical properties such as high stiffness to weight and low density.

In order to use them efficiently, a clear understanding of their behaviors such as: deformable characteristic, stress distribution, natural frequency and critical buckling load under various conditions are required. Hence, investigation on property of composite structure has been addressed since long time. Pagano [1] initially investigated the analytical three-dimensional (3D) elasticity method to predict the exact solution of simple static problems. Noor et al. [2,3] have further developed an 3D elasticity solution formulation for stress analysis of composite structures. It is well known that such an exact 3D approach is the most potential tool to obtain the true solution of plates. However, there is not easy to solve practical problems in which complex (or even slightly complicated) geometries and boundary conditions are required. In addition, each layer in the 3D elasticity theory is modeled as a 3D solid so that the computational cost of laminated composite plate analyses will be increased significantly. Hence, many equivalent single layer (ESL) plate theories with suitable assumptions [4] have been then proposed to transform the 3D problem to 2D one. Among of the ESL plate theories, the Classical Laminate Plate Theory (CLPT) based on the Love-Kirchoff assumptions is first proposed. Due to ignoring the transverse shear deformation, CLPT merely to provide acceptable results for the thin plate. The First Order Shear Deformation Theory (FSDT) based on Reissner [5] and Mindlin [6], which takes into account the shear effect, was therefore developed. In FSDT model, with the linear in-plane displacement assumption through plate thickness, obtained shear strain/stress distributes inaccurately and does not satisfy the traction free boundary conditions at the plate surfaces. The shear correction factors (SCF) are hence required to rectify the unrealistic shear strain energy part. The values of SCF are quite dispersed through each problem and may be difficult to determine [7]. To bypass the limitations of the FSDT, many kind of Higher-Order Shear Deformable Theories (HSDT), which include higher-order terms in the approximation of the displacement field, have then been devised such as Third-Order Shear Deformation Theory (TSDT) [8-12], trigonometric shear deformation theory [13-17], exponential shear deformation theory (ESDT) [18-20], Refined Plate Theory (RPT) and so on. The RPT model was found in Ref [21] of Senthilnathan et al with one variable lower than that of TSDT of Reddy, and then extended by Shimpi et al [22-24], Thai et al [25-26]. It is worth mentioning that the HSDT models provide better results and yield more accurate and stable solutions (e.g. inter-laminar stresses and displacements) [27,28] than the FSDT ones without requirement the SCF. However, the HSDT requires the $C^1$-continuity of generalized displacement field leading to the second-order derivative of the stiffness formulation and it causes the obstacles in the standard finite element formulations. Several $C^0$ continuous elements [29-32] were then proposed or Hermite interpolation function with the $C^1$-continuity was added for just specific approximation of transverse displacement [8]. It may produce extra unknown variables including derivative of deflection $w_{,x}, w_{,y}, w_{,xy}$ [4] leading to increase in the computational cost. In this paper, we show that $C^1$-continuous elements will be easily achieved by IGA without any additional variables.

---

[*] Corresponding author. *Email address*: nxhung@hcmus.edu.vn (H. Nguyen-Xuan)

Isogeometric approach (IGA) has been recently proposed by Hughes et al. [33] to closely link the gap between Computer Aided Design (CAD) and Finite Element Analysis (FEA). The basic idea is that the IGA uses the same non-uniform rational B-Spline (NURBS) functions in describing the exact geometry of problem and constructing finite approximation for analysis. It is well known that NURBS functions provide a flexible way to make refinement, de-refinement, and degree elevation [34]. They enable us to easily achieve the smoothness of arbitrary continuity order in comparison with the traditional FEM. Hence, IGA naturally verifies the $C^1$-continuity of plates based on the HSDT assumption, which is interested in this study. The IGA has been well known and widely applied to various practical problems [35-42] and so on.

In this paper, a combination between Isogeometric Approach and the RPT model (RPT-IGA) for static, free vibration and buckling analysis of laminated composite plates is studied. Herein, some higher-order distributed functions [8,18,19,22,43] are utilized to multiply to higher-order term in displacement field. Several numerical examples are given to show the performance of the proposed method and results obtained are compared to other published methods in the literature.

The paper is outlined as follows. Next section introduces the RPT for composite plates. In section 3, the formulation of plate theory based on IGA is described. The numerical results and discussions are provided in section 4. Finally, this article is closed with some concluding remarks.

## 2. The refined plate theory

### 2.1. Displacement field

To consider the effect of shear deformation directly, the higher-order terms are incorporated into the displacement field. A simple and famous theory for bending plate has also given by JN. Reddy based on TSDT [4]:

$$u(x,y,z) = u_0 + z\beta_x + g(z)(\beta_x + w_{,x})$$
$$v(x,y,z) = v_0 + z\beta_y + g(z)(\beta_y + w_{,y}) \quad , \quad \left(\frac{-h}{2} \leq z \leq \frac{h}{2}\right) \quad (1)$$
$$w(x,y) = w_0$$

where $g(z) = -\dfrac{4z^3}{3h^2}$ and the variables $\mathbf{u}_0 = \{u_0 \ v_0\}^T$, $w_0$ and $\boldsymbol{\beta} = \{\beta_x \ \beta_y\}^T$ are the membrane displacements, the deflection of the mid-plane and the rotations in the *y-z, x-z* planes, respectively. By making additional assumptions given in Eq. (2), Senthilnathan et al [21] proposed the refined plate theory model with one reduced variable.

$$w_0 = w_b + w_s; \quad \beta_n = -\nabla w_b \tag{2}$$

where $w_b$ and $w_s$ are defined as the bending and shear components of deflection. Eq. (1) is taken in the simpler form

$$\begin{aligned} u(x,y,z) &= u_0 - zw_{b,x} + g(z)w_{s,x} \\ v(x,y,z) &= v_0 - zw_{b,y} + g(z)w_{s,y} \\ w(x,y) &= w_b + w_s \end{aligned} \tag{3}$$

The relationship between strains and displacements is described by

$$\boldsymbol{\varepsilon} = [\varepsilon_{xx}\ \varepsilon_{yy}\ \gamma_{xy}]^T = \boldsymbol{\varepsilon}_0 + z\boldsymbol{\kappa}_b + g(z)\boldsymbol{\kappa}_s$$

$$\boldsymbol{\gamma} = \begin{bmatrix} \gamma_{xz} & \gamma_{yz} \end{bmatrix}^T = f'(z)\boldsymbol{\varepsilon}_s \qquad f'(z) = g(z)+1 \tag{4}$$

where

$$\boldsymbol{\varepsilon}_0 = \begin{bmatrix} u_{0,x} \\ v_{0,y} \\ u_{0,y}+v_{0,x} \end{bmatrix}, \boldsymbol{\kappa}_b = -\begin{bmatrix} w_{b,xx} \\ w_{b,yy} \\ 2w_{b,xy} \end{bmatrix}, \boldsymbol{\kappa}_s = \begin{bmatrix} w_{s,xx} \\ w_{s,yy} \\ 2w_{s,xy} \end{bmatrix}, \boldsymbol{\varepsilon}_s = \begin{bmatrix} w_{s,x} \\ w_{s,y} \end{bmatrix} \tag{5}$$

It is seen that $f'(z)=0$ at $z=\pm h/2$. It means that traction-free boundary condition at the top and bottom plate surfaces is automatically satisfied. Based on this condition, many kinds of distributed functions $f(z)$ in forms: third-order polynomials by Reddy [8] and Shimpi [22], exponential function by Karama [18], sinusoidal function by Arya [19] and firth-order polynomial by Nguyen-Xuan [43] are listed in Table 1.

Table 1: The various forms of shape function.

| Model | $f(z)$ | $f'(z)$ |
|---|---|---|
| Reddy [8] | $z - \frac{4}{3}z^3/h^2$ | $1 - 4z^2/h^2$ |
| Shimpi [22] | $\frac{5}{4}z - \frac{5}{3}z^3/h^2$ | $\frac{5}{4}\left(1 - 4z^2/h^2\right)$ |
| Karama [18] | $ze^{-2(z/h)^2}$ | $(1 - \frac{4}{h^2}z^2)e^{-2(z/h)^2}$ |
| Arya [19] | $\sin(\frac{\pi}{h}z)$ | $\frac{\pi}{h}\cos(\frac{\pi}{h}z)$ |
| Nguyen-Xuan [43] | $\frac{7}{8}z - \frac{2}{h^2}z^3 + \frac{2}{h^4}z^5$ | $\frac{7}{8} - \frac{6}{h^2}z^2 + \frac{10}{h^4}z^4$ |

## 2.1. Weak form equations for plate problems

A weak form of the static model for the plates under transverse loading $q_0$ can be briefly expressed as:

$$\int_\Omega \delta\boldsymbol{\varepsilon}^T \mathbf{D}^b \boldsymbol{\varepsilon} d\Omega + \int_\Omega \delta\boldsymbol{\gamma}^T \mathbf{D}^s \boldsymbol{\gamma} d\Omega = \int_\Omega \delta w q_0 d\Omega \tag{6}$$

where

$$\mathbf{D}^b = \begin{bmatrix} \mathbf{A} & \mathbf{B} & \mathbf{E} \\ \mathbf{B} & \mathbf{D} & \mathbf{F} \\ \mathbf{E} & \mathbf{F} & \mathbf{H} \end{bmatrix} \tag{7}$$

and the material matrices are given as

$$A_{ij}, B_{ij}, D_{ij}, E_{ij}, F_{ij}, H_{ij} = \int_{-h/2}^{h/2} (1, z, z^2, g(z), zg(z), g^2(z)) \bar{Q}_{ij} dz \quad (i, j = 1, 2, 6)$$

$$D_{ij}^s = \int_{-h/2}^{h/2} [f'(z)]^2 \bar{Q}_{ij} dz \quad (i, j = 4, 5) \tag{8}$$

in which $\bar{Q}_{ij}$ are transformed material constants of the $k^{th}$ lamina (see [4] for more detail).

For the free vibration analysis of the plates, weak form can be derived from the following dynamic equation

$$\int_\Omega \delta\boldsymbol{\varepsilon}^T \mathbf{D}^b \boldsymbol{\varepsilon} d\Omega + \int_\Omega \delta\boldsymbol{\gamma}^T \mathbf{D}^s \boldsymbol{\gamma} d\Omega = \int_\Omega \delta\tilde{\mathbf{u}}^T \mathbf{m} \ddot{\tilde{\mathbf{u}}} d\Omega \tag{9}$$

where $\mathbf{m}$ - the mass matrix is calculated according to the consistent form

$$\mathbf{m} = \begin{bmatrix} \mathbf{I_0} & 0 & 0 \\ 0 & \mathbf{I_0} & 0 \\ 0 & 0 & \mathbf{I_0} \end{bmatrix} \text{ where } \mathbf{I_0} = \begin{bmatrix} I_1 & I_2 & I_4 \\ I_2 & I_3 & I_5 \\ I_4 & I_5 & I_6 \end{bmatrix} \tag{10}$$

$$(I_1, I_2, I_3, I_4, I_5, I_6) = \int_{-h/2}^{h/2} \rho(z)(1, z, z^2, g(z), zg(z), g^2(z)) dz . \tag{11}$$

and

$$\tilde{\mathbf{u}} = \begin{Bmatrix} \mathbf{u}_1 \\ \mathbf{u}_2 \\ \mathbf{u}_3 \end{Bmatrix}, \quad \mathbf{u}_1 = \begin{Bmatrix} u_0 \\ -w_{b,x} \\ w_{s,x} \end{Bmatrix}; \mathbf{u}_2 = \begin{Bmatrix} v_0 \\ -w_{b,y} \\ w_{s,y} \end{Bmatrix}; \mathbf{u}_3 = \begin{Bmatrix} w \\ 0 \\ 0 \end{Bmatrix} \tag{12}$$

For the buckling analysis, a weak form of the plate under the in-plane forces can be expressed as:

$$\int_\Omega \delta\boldsymbol{\varepsilon}^T \mathbf{D}^b \boldsymbol{\varepsilon} \mathrm{d}\Omega + \int_\Omega \delta\boldsymbol{\gamma}^T \mathbf{D}^s \boldsymbol{\gamma} \mathrm{d}\Omega + \int_\Omega \nabla^T \delta w \mathbf{N}_0 \nabla w \mathrm{d}\Omega = 0 \tag{13}$$

where $\nabla^T = [\partial/\partial x \; \partial/\partial y]^T$ is the gradient operator and $\mathbf{N}_0 = \begin{bmatrix} N_x^0 & N_{xy}^0 \\ N_{xy}^0 & N_y^0 \end{bmatrix}$ is a matrix related to the pre-buckling loads.

## 3. The composite plate formulation based on NURBS basis functions

*3.1. A brief of NURBS functions*

A knot vector $\Xi = \{\xi_1, \xi_2, ..., \xi_{n+p+1}\}$ is defined as a sequence of knot value $\xi_i \in R$, $i = 1,...n+p$. If the first and the last knots are repeated $p+1$ times, the knot vector is called open knot. A B-spline basis function is $C^\infty$ continuous inside a knot span and $C^{p-1}$ continuous at a single knot. Thus, as $p \geq 2$ the present approach always satisfies $C^1$-requirement in approximate formulations of RPT.

The B-spline basis functions $N_{i,p}(\xi)$ are defined by the following recursion formula

$$N_{i,p}(\xi) = \frac{\xi - \xi_i}{\xi_{i+p} - \xi_i} N_{i,p-1}(\xi) + \frac{\xi_{i+p+1} - \xi}{\xi_{i+p+1} - \xi_{i+1}} N_{i+1,p-1}(\xi)$$

$$\text{as } p = 0, \; N_{i,0}(\xi) = \begin{cases} 1 & \text{if } \xi_i < \xi < \xi_{i+1} \\ 0 & \text{otherwise} \end{cases} \tag{14}$$

By the tensor product of basis functions in two parametric dimensions $\xi$ and $\eta$ with two knot vectors $\Xi = \{\xi_1, \xi_2, ..., \xi_{n+p+1}\}$ and $\mathbf{H} = \{\eta_1, \eta_2, ..., \eta_{m+q+1}\}$, the two-dimensional B-spline basis functions are obtained

$$N_A(\xi, \eta) = N_{i,p}(\xi) M_{j,q}(\eta) \tag{15}$$

Figure 1 illustrates the set of one-dimensional and two-dimensional B-spline basis functions according to open uniform knot vector $\Xi = \{0, 0, 0, 0, 0.5, 1, 1, 1, 1\}$.

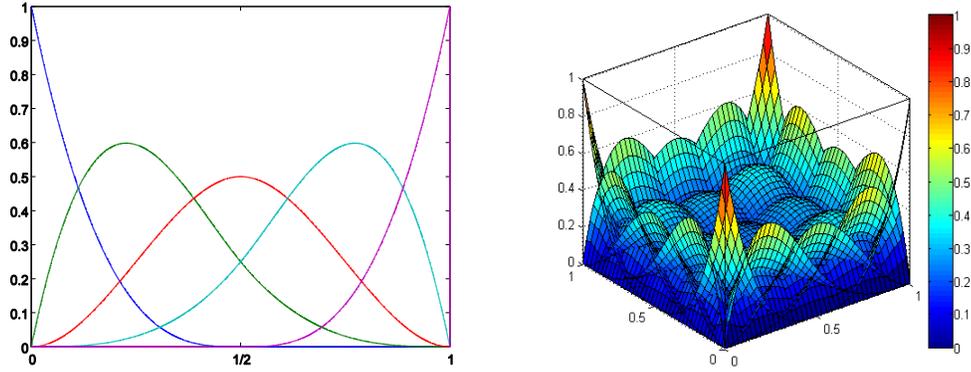

Figure 1. 1D and 2D B-spline basis functions.

To present exactly some curved geometries (e.g. circles, cylinders, spheres, etc.) the non-uniform rational B-splines (NURBS) functions are used. Be different from B-spline, each control point of NURBS has additional value called an individual weight $w_A$ [33]. Then the NURBS functions can be expressed as

$$R_A(\xi,\eta) = \frac{N_A w_A}{\sum_{A}^{m \times n} N_A(\xi,\eta) w_A} \tag{16}$$

It can be noted that B-spline function is only the special case of the NURBS function when the individual weight of control point is constant.

*3.2. A novel RPT formulation based on NURBS approximation*

Using the NURBS basis functions above, the displacement field **u** of the plate is approximated as

$$\mathbf{u}^h(\xi,\eta) = \sum_{A}^{m \times n} R_A(\xi,\eta) \mathbf{q}_A \tag{17}$$

where $\mathbf{q}_A = [u_{0A}\ v_{0A}\ w_{bA}\ w_{sA}]^T$ is the vector of nodal degrees of freedom associated with the control point A.

Substituting Eq. (17) into Eq. (5), the in-plane and shear strains can be rewritten as:

$$\left[\boldsymbol{\varepsilon}_0^T\ \boldsymbol{\kappa}_b^T\ \boldsymbol{\kappa}_s^T\ \boldsymbol{\varepsilon}_s^T\right]^T = \sum_{A=1}^{m \times n}\left[\left(\mathbf{B}_A^m\right)^T\ \left(\mathbf{B}_A^{b1}\right)^T\ \left(\mathbf{B}_A^{b2}\right)^T\ \left(\mathbf{B}_A^s\right)^T\right]^T \mathbf{q}_A \tag{18}$$

in which

$$\mathbf{B}_A^m = \begin{bmatrix} R_{A,x} & 0 & 0 & 0 \\ 0 & R_{A,y} & 0 & 0 \\ R_{A,y} & R_{A,x} & 0 & 0 \end{bmatrix}, \quad \mathbf{B}_A^{b1} = -\begin{bmatrix} 0 & 0 & R_{A,xx} & 0 \\ 0 & 0 & R_{A,yy} & 0 \\ 0 & 0 & 2R_{A,xy} & 0 \end{bmatrix},$$

$$\mathbf{B}_A^{b2} = \begin{bmatrix} 0 & 0 & 0 & R_{A,xx} \\ 0 & 0 & 0 & R_{A,yy} \\ 0 & 0 & 0 & 2R_{A,xy} \end{bmatrix}, \quad \mathbf{B}_A^s = \begin{bmatrix} 0 & 0 & 0 & R_{A,x} \\ 0 & 0 & 0 & R_{A,y} \end{bmatrix}$$

(19)

Substituting Eq. (18) into Eqs. (6), (9) and (13), the formulations of static, free vibration and buckling problem are rewritten in the following form

$$\mathbf{K}\mathbf{q} = \mathbf{F} \tag{20}$$

$$\left(\mathbf{K} - \omega^2 \mathbf{M}\right)\mathbf{q} = \mathbf{0} \tag{21}$$

$$\left(\mathbf{K} - \lambda_{cr} \mathbf{K}_g\right)\mathbf{q} = \mathbf{0} \tag{22}$$

where the global stiffness matrix $\mathbf{K}$ is given by

$$\mathbf{K} = \int_\Omega \begin{Bmatrix} \mathbf{B}^m \\ \mathbf{B}^{b1} \\ \mathbf{B}^{b2} \end{Bmatrix}^T \begin{bmatrix} \mathbf{A} & \mathbf{B} & \mathbf{E} \\ \mathbf{B} & \mathbf{D} & \mathbf{F} \\ \mathbf{E} & \mathbf{F} & \mathbf{H} \end{bmatrix} \begin{Bmatrix} \mathbf{B}^m \\ \mathbf{B}^{b1} \\ \mathbf{B}^{b2} \end{Bmatrix} + \mathbf{B}^{sT} \mathbf{D}^s \mathbf{B}^s \, \mathrm{d}\Omega \tag{23}$$

and the load vector is computed by

$$\mathbf{F} = \int_\Omega q_0 \mathbf{R} \, \mathrm{d}\Omega \tag{24}$$

where

$$\mathbf{R} = \begin{bmatrix} 0 & 0 & R_A & R_A \end{bmatrix} \tag{25}$$

the global mass matrix $\mathbf{M}$ is expressed as

$$\mathbf{M} = \int_\Omega \tilde{\mathbf{R}}^T \mathbf{m} \tilde{\mathbf{R}} \, \mathrm{d}\Omega \tag{26}$$

where

$$\tilde{\mathbf{R}} = \begin{Bmatrix} \mathbf{R}_1 \\ \mathbf{R}_2 \\ \mathbf{R}_3 \end{Bmatrix}, \quad \mathbf{R}_1 = \begin{bmatrix} R_A & 0 & 0 & 0 \\ 0 & 0 & -R_{A,x} & 0 \\ 0 & 0 & 0 & R_{A,x} \end{bmatrix};$$

$$\mathbf{R}_2 = \begin{bmatrix} 0 & R_A & 0 & 0 \\ 0 & 0 & -R_{A,y} & 0 \\ 0 & 0 & 0 & R_{A,y} \end{bmatrix}; \mathbf{R}_3 = \begin{bmatrix} 0 & 0 & R_A & R_A \\ 0 & 0 & 0 & 0 \\ 0 & 0 & 0 & 0 \end{bmatrix}$$

(27)

the geometric stiffness matrix is

$$\mathbf{K}_g = \int_\Omega \left(\mathbf{B}^g\right)^T \mathbf{N}_0 \mathbf{B}^g \, d\Omega \tag{28}$$

where

$$\mathbf{B}_A^g = \begin{bmatrix} 0 & 0 & R_{A,x} & R_{A,x} \\ 0 & 0 & R_{A,y} & R_{A,y} \end{bmatrix} \tag{29}$$

in which $\omega$, $\lambda_{cr} \in R^+$ are the natural frequency and the critical buckling value, respectively.

It is observed from Eq. (23) that the SCF is no longer required in the stiffness formulation. Herein, $\mathbf{B}_A^{b1}$ and $\mathbf{B}_A^{b2}$ contain the second-order derivative of the shape functions. Hence, it requires $C^1$-continuous element in approximate formulations. It is now interesting to note that our present formulation based on IGA naturally satisfies $C^1$-continuity from the theoretical/mechanical viewpoint of plates [40, 28]. In our work, the basis functions are $C^{p-1}$ continuous. Therefore, as $p \geq 2$, the present approach always satisfies $C^1$-requirement in approximate formulations based on the proposed RPT.

*3.2. Essential boundary conditions*

Herein, many kinds of boundary condition are applied for an arbitrary edge with simply supported (S) and clamped (C) conditions including:

Simply supported cross-ply:

$$\begin{aligned} v_0 = w_b = w_s = 0 \quad &\text{at} \ x = 0, a \\ u_0 = w_b = w_s = 0 \quad &\text{at} \ y = 0, b \end{aligned} \tag{30}$$

Simply supported angle-ply:

$$\begin{aligned} u_0 = w_b = w_s = 0 \quad &\text{at} \ x = 0, a \\ v_0 = w_b = w_s = 0 \quad &\text{at} \ y = 0, b \end{aligned} \tag{31}$$

Clamped (C):

$$u_0 = v_0 = w_b = w_s = w_{b,n} = w_{s,n} = 0 \tag{32}$$

The Dirichlet boundary condition on $u_0, v_0, w_b$ and $w_s$ is easily treated as in the standard FEM. However, for the derivatives $w_{b,n}, w_{s,n}$ the enforcement of Dirichlet BCs can be solved in a simple and effective way [44]. The idea is as follows. The derivatives can be included in a compact form of the normal slope at the boundary:

$$\frac{\partial w}{\partial n} = \lim_{\Delta n \to 0} \frac{w(n(\mathbf{x}_C) + \Delta n) - w(n(\mathbf{x}_C))}{\Delta n} = 0 \tag{33}$$

As $w(n(\mathbf{x}_C)) = 0$ according to Eq. (32), Eq. (33) leads to impose the same boundary values, i.e, zero values, on the deflection variable at control points $\mathbf{x}_A$ which is adjacent to the boundary control points $\mathbf{x}_C$. It can be observed that, implementing the essential boundary condition using this method is very simple in IGA compare to other numerical methods.

## 4. Results and discussions

In this section, we show the performance of the present method – RPT-IGA with various distributed functions given in Table 1 in analyzing the laminated composite plates. We illustrate the method using the cubic basis functions with full $(p+1) \times (q+1)$ Gauss points.

*4.1 Static analysis*

In this sub-section, material set I is used with parameters given as:

Material I:

$E_1 = 25E_2, G_{12} = G_{13} = 0.5E_2, G_{23} = 0.2E_2, \nu_{12} = 0.25, \rho = 1.$

For convenience, the following normalized transverse displacement, in-plane stresses and shear stresses are expressed as:

$$\bar{w} = \frac{10^2 w E_2 h^3}{q_0 a^4}, \bar{\sigma} = \frac{\sigma h^2}{q_0 a^2}, \bar{\tau} = \frac{\tau h}{q_0 a}$$

Let us consider the simply supported square plate shown in Figure 2a. The plates stacked by two layer [0/90] subjected to a sinusoidal pressure defined as $q_0 \sin(\frac{\pi x}{a})\sin(\frac{\pi y}{a})$ at the top surface.

We first investigate the convergence of the normalization displacement and stresses with length to thickness ratio $a/h$ =10. The plate is modeled with 7×7, 11×11 and 15×15 cubic elements as shown in Figure 2. The comparisons between present results using RPT and analytical solution using HSDT given by Khdeir and Reddy [47] are tabulated in Table 2. The relative error is given in the parentheses. It can be seen that the obtained results agree well with the exact values. It is observed that model using third-order polynomials gains the most accuracy displacement while that using fifth-order one (FiSDT) obtains the closest axial stress. IGA, moreover, gains high-convergence with coarse mesh using 11×11 cubic elements. Therefore, in the next problems, the meshing of 11x11 cubic NURBS elements shown in Figure 2c is used.

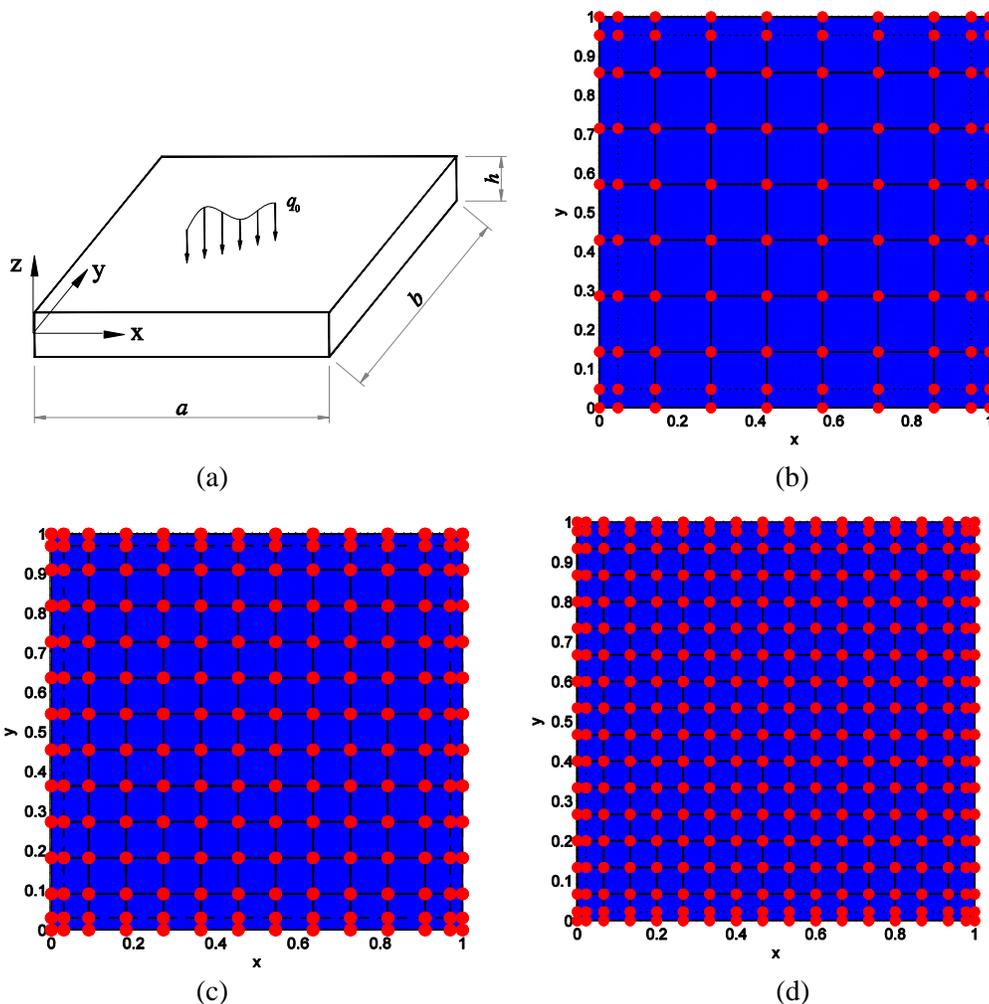

Figure 2. Square plate: (a) The plate geometry; (b), (c), (d): meshing of 7x7, 11x11, 15x15 cubic elements, respectively.

Table 2: The convergence of the normalized displacement and stresses of the simply supported [0/90] composite plate (*a/h* =10) under sinusoidal load.

| Plate model | | Nor. Sol. | Mesh | | |
|---|---|---|---|---|---|
| | | | 7x7 | 11x11 | 15x15 |
| RPT-IGA | Reddy [8] | $\bar{w}(\frac{a}{2},\frac{b}{2})$ | 1.2159 (0.01) | 1.2161 (-0.01) | 1.2161 (-0.01) |
| | Shimpi [22] | | 1.2159 (0.01) | 1.2161 (-0.01) | 1.2161 (-0.01) |
| | Arya [19] | | 1.2129 (0.25) | 1.2131 (0.24) | 1.2131 (0.24) |
| | Karama [18] | | 1.2093 (0.55) | 1.2096 (0.53) | 1.2096 (0.53) |
| | FiSDT [43] | | 1.2041 (0.98) | 1.2043 (0.96) | 1.2043 (0.96) |
| **Analytical Solution [47]** | | | | | **1.216** |
| RPT-IGA | Reddy [8] | $\bar{\sigma}_x(\frac{a}{2},\frac{b}{2},-\frac{h}{2})$ | -0.7351 (1.57) | -0.7421 (0.63) | -0.7443 (0.33) |
| | Shimpi [22] | | -0.7351 (1.57) | -0.7421 (0.63) | -0.7443 (0.33) |
| | Arya [19] | | -0.7366 (1.37) | -0.7436 (0.43) | -0.7458 (0.13) |
| | Karama [18] | | -0.7379 (1.19) | -0.7449 (0.25) | -0.7471 (-0.04) |
| | FiSDT [43] | | -0.7397 (0.95) | -0.7467 (0.01) | -0.7489 (-0.28) |
| **Analytical Solution [47]** | | | | | **-0.7468** |

(*) The error in parentheses

Next, the behavior of two-layer [0/90] square composite plate under two types of boundary condition (SSSS and SFSF) are considered; here, S = simply supported and F = free edge. The obtained results of present model are compared with those published one from 3D model of Vel and Batra [46]; HSDT, FSDT, CLPT using analytical solution by Khdeir and Reddy [47] and HOSNDPT using mesh free with 18DOFs/node by Xiao et al [45]. The comparison is provided in Table 3. The normalized displacement and stresses of the present approach are in good agreement with the 3D exact solution [46]. Among present models, the third-order distributed functions by Reddy [8] and Shimpi [22] combined in RPT-IGA archives the same results which are closest to 3D exact solution. The transverse displacement of plates is illustrated in Figure 3 according to SFSF and SSSS boundary conditions, respectively. Figure 4 plots the stress distribution through the thickness of composite plate under full simply supported condition with *a/h* = 10. Herein, the NURBS functions are used to model the RPT assumption with various *f(z)* functions such as: TSDT of Reddy [8], HSDT of Shimpi [22], ESDT of Karama [18], SSDT of Arya [19] and FiSDT of Nguyen-Xuan [43]. Using RPT model, the in-plane stresses is plotted in the same path while there is a slight difference observed for shear stress distribution. And, all of them satisfy the traction-free boundary conditions at the plate surfaces.

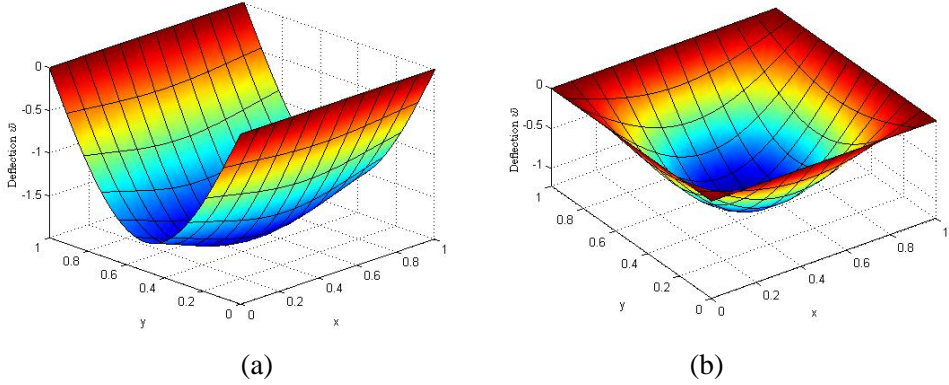

(a) (b)
Figure 3. Deflection profile of Al/ZrO$_2$-1 FGM plates: (a) SFSF; (b) SSSS.

Table 3: The non-dimensional deflection and stresses of two-layer [0/90] square composite plate under sinusoidal load.

| BC | Plate model | | $\bar{w}(\frac{a}{2},\frac{b}{2})$ | $\bar{\sigma}_x(\frac{a}{2},\frac{b}{2},-\frac{h}{2})$ | $\bar{\sigma}_y(\frac{a}{2},\frac{b}{2},\frac{h}{2})$ | $\bar{\sigma}_{xy}(0,0,\frac{h}{2})$ | $\bar{\sigma}_{yz}(\frac{a}{2},0,0)$ |
|---|---|---|---|---|---|---|---|
| **SSSS** | **3D model [46]** | | **1.227** | **-0.7304** | **0.7309** | **0.0497** | **-** |
| | HSDT[47] | Exact | 1.216 | -0.7468 | 0.7468 | - | 0.319 |
| | | FEM | 1.214 | -0.6829 | 0.6829 | - | - |
| | FSDT[47] | | 1.237 | -0.7157 | 0.7157 | - | 0.2729 |
| | CLPT [47] | | 1.064 | -0.7157 | 0.7157 | - | 0 |
| | HOSNDPT [45] | MQ-MLPG | 1.22 | -0.726 | 0.727 | 0.0494 | 0.298 |
| | | TPS-MLPG | 1.213 | -0.723 | 0.724 | 0.0491 | 0.278 |
| | RPT-IGA | Reddy [8] | 1.2161 | -0.7421 | 0.7421 | 0.053 | 0.3181 |
| | | Shimpi [22] | 1.2161 | -0.7421 | 0.7421 | 0.053 | 0.3181 |
| | | Arya [19] | 1.2131 | -0.7436 | 0.7436 | 0.053 | 0.3252 |
| | | Karama [18] | 1.2096 | -0.7449 | 0.7449 | 0.0531 | 0.3319 |
| | | FiSDT [43] | 1.2043 | -0.7467 | 0.7467 | 0.0531 | 0.3369 |
| **SFSF** | **3D model [46]** | | **2.026** | **0.2503** | **1.210** | **0.0119** | **-** |
| | HSDT[47] | Exact | 1.992 | 0.2624 | 1.2295 | - | 0.4489 |
| | | FEM | 2.002 | 0.2212 | 1.189 | - | - |
| | FSDT[47] | | 2.028 | 0.2469 | 1.1907 | - | 0.3882 |
| | CLPT [47] | | 1.777 | 0.2403 | 1.1849 | - | 0 |
| | HOSNDPT | MQ-MLPG | 2.028 | 0.249 | 1.21 | 0.0118 | 0.488 |

|  |  |  |  |  |  |  |
|---|---|---|---|---|---|---|
| [45] | TPS-MLPG | 2.028 | 0.249 | 1.21 | 0.0119 | 0.499 |
| RPT-IGA | Reddy [8] | 1.990 | 0.25472 | 1.2192 | 0.0121 | 0.4507 |
|  | Shimpi [22] | 1.990 | 0.25472 | 1.2192 | 0.0121 | 0.4507 |
|  | Arya [19] | 1.9851 | 0.2555 | 1.221 | 0.0121 | 0.46 |
|  | Karama [18] | 1.9794 | 0.2562 | 1.2225 | 0.0122 | 0.4686 |
|  | FiSDT [43] | 1.9712 | 0.2572 | 1.2247 | 0.0122 | 0.4744 |

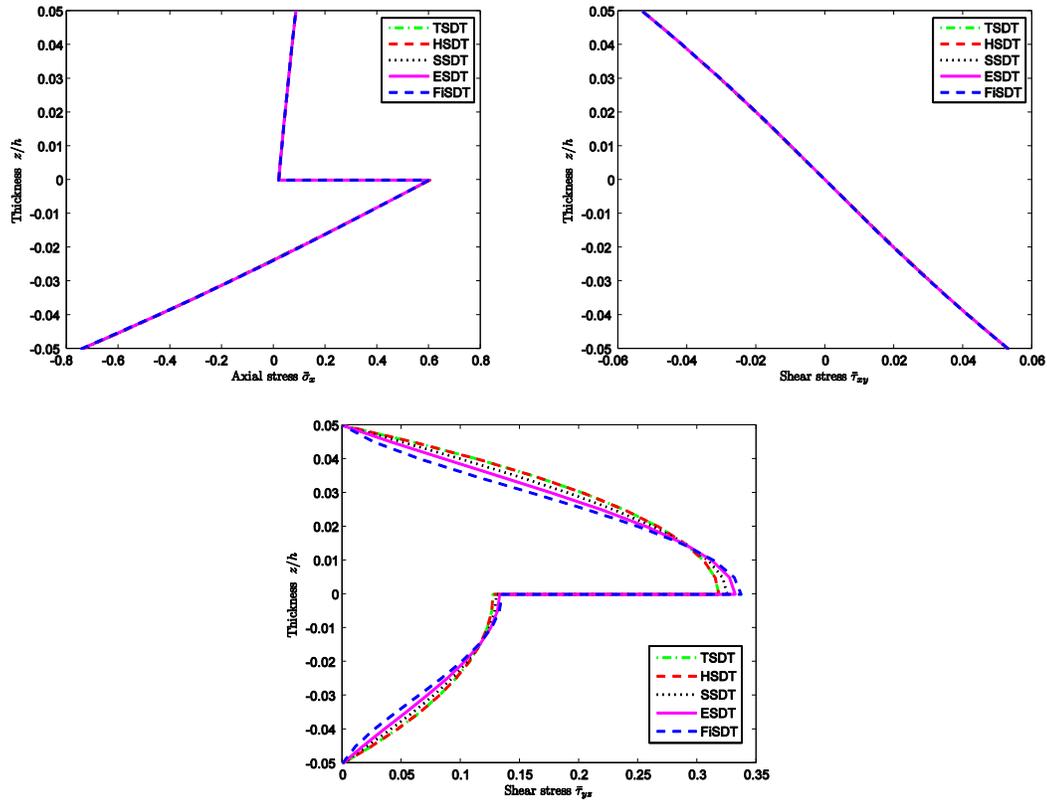

Figure 4. The stresses through thickness of laminate composite plate under full simply supported condition with *a/h*=10 via different refined plate models.

The relation between non-dimensional deflection and length to thickness ratio is depicted in Table 4. It is observed that the present results using isogeometric finite element approach gain the identical solutions compared to PRT [26] and TSDT [49] one using analytical solution, especially that of Reddy's and Shimpi's models. Moreover, as plate becomes thinner (*a/h* increases) the difference between those solutions is not significant.

Table 4: The non-dimensional deflection of simply supported two-layer (0/90) square plate under sinusoidal load.

| Plate model | | a/h | | | |
|---|---|---|---|---|---|
| | | 5 | 10 | 20 | 100 |
| 3D model [1] | | **1.7287** | **1.2318** | **1.106** | **1.0742** |
| TSDT [49] | | 1.667 | 1.2161 | 1.1018 | 1.0651 |
| FSDT [48] | | 1.667 | 1.2416 | 1.1113 | 1.0653 |
| RPT - analytical [26] | | 1.667 | 1.2161 | 1.1018 | 1.0651 |
| RPT-IGA | Reddy [8] | 1.6669 | 1.2161 | 1.1018 | 1.0651 |
| | Shimpi [22] | 1.6669 | 1.2161 | 1.1018 | 1.0651 |
| | Arya [19] | 1.6538 | 1.2131 | 1.1011 | 1.065 |
| | Karama [18] | 1.6382 | 1.2096 | 1.1002 | 1.065 |
| | FiSDT [43] | 1.6154 | 1.2043 | 1.0989 | 1.065 |

*4.2 Free vibration analysis*

Let us consider the cross-ply $[0/90]_N$ composite plate with $a/h=5$ and simply supported boundary conditions. Herein, used material set II is defined as:

$E_1/E_2 = open$, $G_{12} = G_{13} = 0.6E_2$, $G_{23} = 0.5E_2$, $\nu_{12} = 0.25$, $\rho = 1$.

The effects of the number of layers *N* and elastic modulus ratios $E_1/E_2$ are tabulated in Table 5. A good agreement is found for present models in comparison with three-dimensional elastic solutions proposed by Noor [2] and the analytical solution given by Kant [50]. It is depicted that, among present model, FiSDT archives the highest results which is closest to 3D solution as $E_1/E_2 \leq 10$. As $E_1/E_2$ ratio ranges from 20 to 40, the present results are asymptotic to analytical solutions for 2D plate model using HSDT and RPT [50], especially, the model using third-order functions.

Table 5: The natural frequency $\bar{\omega} = \omega a^2 / h \sqrt{\rho / E_2}$ of simply supported $[0/90]_N$ composite plate.

| N | Model | | $E_1/E_2$ | | | | |
|---|---|---|---|---|---|---|---|
| | | | 3 | 10 | 20 | 30 | 40 |
| 1 | 3D elasticity [2] | | 6.2578 | 6.9845 | 7.6745 | 8.1763 | 8.5625 |
| | Anal. Sol. [50] | HSDT-12DOFs | 6.2336 | 6.9741 | 7.714 | 8.2775 | 8.7272 |
| | | HSDT-9DOFs | 6.1566 | 6.9363 | 7.6883 | 8.257 | 8.7097 |
| | | RPT | 6.2169 | 6.9887 | 7.821 | 8.505 | 9.0871 |
| | | FSDT | 6.149 | 6.9156 | 7.6922 | 8.3112 | 8.8255 |
| | RPT-IGA | Reddy [8] | 6.2169 | 6.9887 | 7.8211 | 8.5051 | 9.0872 |
| | | Shimpi [22] | 6.2169 | 6.9887 | 7.8211 | 8.5051 | 9.0872 |
| | | Arya [19] | 6.2189 | 6.9965 | 7.838 | 8.5317 | 9.1237 |
| | | Karama [18] | 6.2224 | 7.0066 | 7.8585 | 8.563 | 9.1662 |
| | | FiSDT [43] | 6.2296 | 7.0231 | 7.8892 | 8.6089 | 9.2275 |
| 2 | 3D elasticity [2] | | 6.5455 | 8.1445 | 9.4055 | 10.165 | 10.6798 |
| | Anal. Sol. [50] | HSDT-12DOFs | 6.5146 | 8.1482 | 9.4675 | 10.2733 | 10.8221 |
| | | HSDT-9DOFs | 6.4319 | 8.101 | 9.4338 | 10.2463 | 10.7993 |
| | | TSDT, RPT | 6.5008 | 9.1954 | 9.6265 | 10.5348 | 11.1716 |
| | | FSDT | 6.4402 | 8.1963 | 9.6729 | 10.6095 | 11.2635 |
| | | Reddy [8] | 6.5008 | 8.1954 | 9.6265 | 10.5348 | 11.1716 |
| | | Shimpi [22] | 6.5008 | 8.1954 | 9.6265 | 10.5348 | 11.1716 |
| | RPT-IGA | Arya [19] | 6.5012 | 8.193 | 9.6205 | 10.5268 | 11.1628 |
| | | Karama [18] | 6.5034 | 8.1939 | 9.6201 | 10.5261 | 11.1629 |
| | | FiSDT [43] | 6.5094 | 8.2021 | 9.6316 | 10.5418 | 11.1832 |
| 3 | 3D elasticity [2] | | 6.61 | 8.4143 | 9.8398 | 10.6958 | 11.2728 |
| | | HSDT-12DOFs | 6.5711 | 8.3852 | 9.8346 | 10.7113 | 11.3051 |
| | Anal. | HSDT-9DOFs | 6.4873 | 8.3372 | 9.8012 | 10.6853 | 11.2838 |
| | Sol. [50] | TSDT, RPT | 6.5552 | 8.4041 | 9.9175 | 10.8542 | 11.5007 |
| | | FSDT | 6.4916 | 8.3883 | 9.9266 | 10.8723 | 11.5189 |
| | | Reddy [8] | 6.5558 | 8.4052 | 9.9181 | 10.8547 | 11.5012 |
| | | Shimpi [22] | 6.5558 | 8.4052 | 9.9181 | 10.8547 | 11.5012 |
| | RPT-IGA | Arya [19] | 6.5567 | 8.4066 | 9.9211 | 10.8604 | 11.5103 |
| | | Karama [18] | 6.5596 | 8.4122 | 9.9313 | 10.8758 | 11.5314 |
| | | FiSDT [43] | 6.5663 | 8.4259 | 9.9555 | 10.9106 | 11.5768 |
| 5 | 3D elasticity [2] | | 6.6458 | 8.5625 | 10.0843 | 11.0027 | 11.6245 |
| | | HSDT-12DOFs | 6.6019 | 8.5163 | 10.0438 | 10.9699 | 11.5993 |
| | Anal. | HSDT-9DOFs | 6.5177 | 8.468 | 10.0107 | 10.9445 | 11.5789 |
| | Sol. [50] | TSDT, RPT | 6.5842 | 8.5126 | 10.0674 | 11.0197 | 11.673 |
| | | FSDT | 6.5185 | 8.4842 | 10.0483 | 10.9959 | 11.6374 |
| | | Reddy [8] | 6.5842 | 8.5126 | 10.0674 | 11.0197 | 11.673 |
| | | Shimpi [22] | 6.5842 | 8.5126 | 10.0674 | 11.0197 | 11.673 |
| | RPT-IGA | Arya [19] | 6.5854 | 8.5156 | 10.0741 | 11.031 | 11.6894 |
| | | Karama [18] | 6.5885 | 8.5229 | 10.0882 | 11.0523 | 11.7182 |
| | | FiSDT [43] | 6.5957 | 8.5394 | 10.1185 | 11.0964 | 11.7757 |

Next, with constant $E_1/E_2$ ratio (equal 40), the variation of natural frequency of two-layer laminate composite plate via length to thickness ratio are listed in Table 6. It is again seen that the obtained results match well with analytical one using 12DOFs published by Kant [50]. The difference reduces via the increase in the ratio of $a/h$ (from approximate 8% to 0.02% according to $a/h$ = 4 and 100, respectively). The first three mode shapes of thick plate ($a/h$=10) is then plotted in Figure 5. It is clear that beside the full mode shape of deflection (above), the mode shapes of all unknown parameters along line $y=a/2$ are illustrated below.

Table 6: The natural frequency $\bar{\omega}=\omega a^2/h\sqrt{\rho/E_2}$ of simply supported [0/90] composite plate with $E_1/E_2 = 40$.

| Plate model | | a/h | | | | |
|---|---|---|---|---|---|---|
| | | 4 | 10 | 20 | 50 | 100 |
| | HSDT-12DOFs | 7.9081 | 10.4319 | 11.0663 | 11.2688 | 11.2988 |
| | HSDT-9DOFs | 7.8904 | 10.4156 | 11.0509 | 11.2537 | 11.2837 |
| Anal. | TSDT | 8.3546 | 10.568 | 11.1052 | 11.2751 | 11.3002 |
| Sol [50] | RPT | 8.3546 | 10.568 | 11.1052 | 11.2751 | 11.3002 |
| | FSDT | 8.0889 | 10.461 | 11.0639 | 11.2558 | 11.2842 |
| RPT-IGA | Reddy [8] | 8.3547 | 10.5681 | 11.1053 | 11.2752 | 11.3003 |
| | Shimpi [22] | 8.3547 | 10.5681 | 11.1053 | 11.2752 | 11.3003 |
| | Arya [19] | 8.4018 | 10.5812 | 11.109 | 11.2758 | 11.3004 |
| | Karama [18] | 8.4564 | 10.5965 | 11.1133 | 11.2758 | 11.3006 |
| | FiSDT [43] | 8.5355 | 10.6186 | 11.1196 | 11.2776 | 11.3009 |

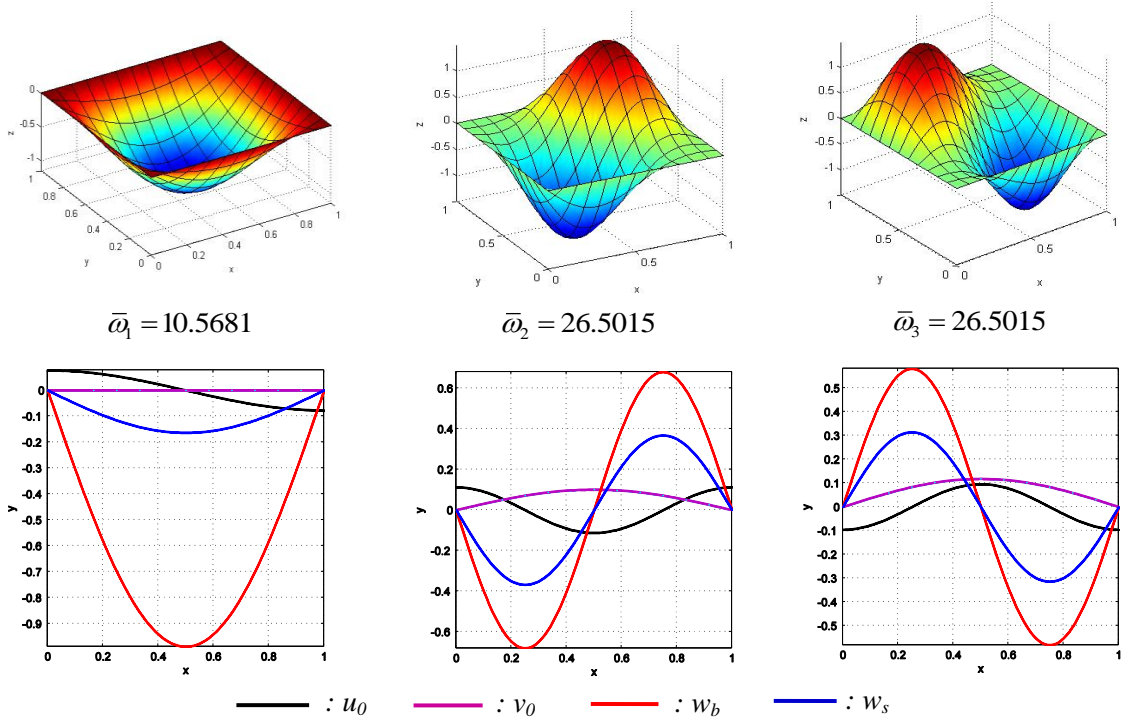

Figure 5. Vibration mode shapes: full plate (upper) and line $y = a/2$ (lower) of simply supported [0/90] composite plate with $E_1/E_2 = 40$, $a/h=10$.

To close this sub-section, the effect of boundary condition on normalized frequency of ten-layer cross-ply composite plate is investigated in Table 7. Compared with those reported by Reddy and Khdeir [51], the present models gain the good agreement. It can be seen that, present models using RPT gain the closest results to analytical solution using TSDT with slightly higher results. In addition, when the constrained edge changes from F to S and C, the structural stiffness increases, the magnitudes of free vibration thus increase, respectively. The mode shapes according to various boundary conditions are illustrated in Figure 6.

Table 7: The natural frequency of ten-layer $[0/90]_5$ composite plate with $a/h=5$ and $E_1/E_2 = 40$

| Plate model | Boundary conditions |
|---|---|

|  |  | SFSF | SFSC | SSSS | SSSC | SCSC | CCCC |
|---|---|---|---|---|---|---|---|
| Anal. Sol. [51] | TSDT | 8.155 | 8.966 | 11.673 | 12.514 | 13.568 | - |
|  | FSDT | 8.139 | 8.919 | 11.644 | 12.197 | 12.923 | - |
|  | CLPT | 11.459 | 13.618 | 12.167 | 23.348 | 30.855 | - |
| RPT-IGA | Reddy [8] | 8.1554 | 9.0832 | 11.673 | 13.0041 | 14.1566 | 15.2991 |
|  | Shimpi [22] | 8.1554 | 9.0832 | 11.673 | 13.0041 | 14.1566 | 15.2991 |
|  | Arya [19] | 8.1661 | 9.0971 | 11.6894 | 13.0463 | 14.2418 | 15.4558 |
|  | Karama [18] | 8.1853 | 9.1201 | 11.7182 | 13.1062 | 14.3513 | 15.6438 |
|  | FiSDT [43] | 8.2238 | 9.1646 | 11.7757 | 13.2122 | 14.5318 | 15.9367 |

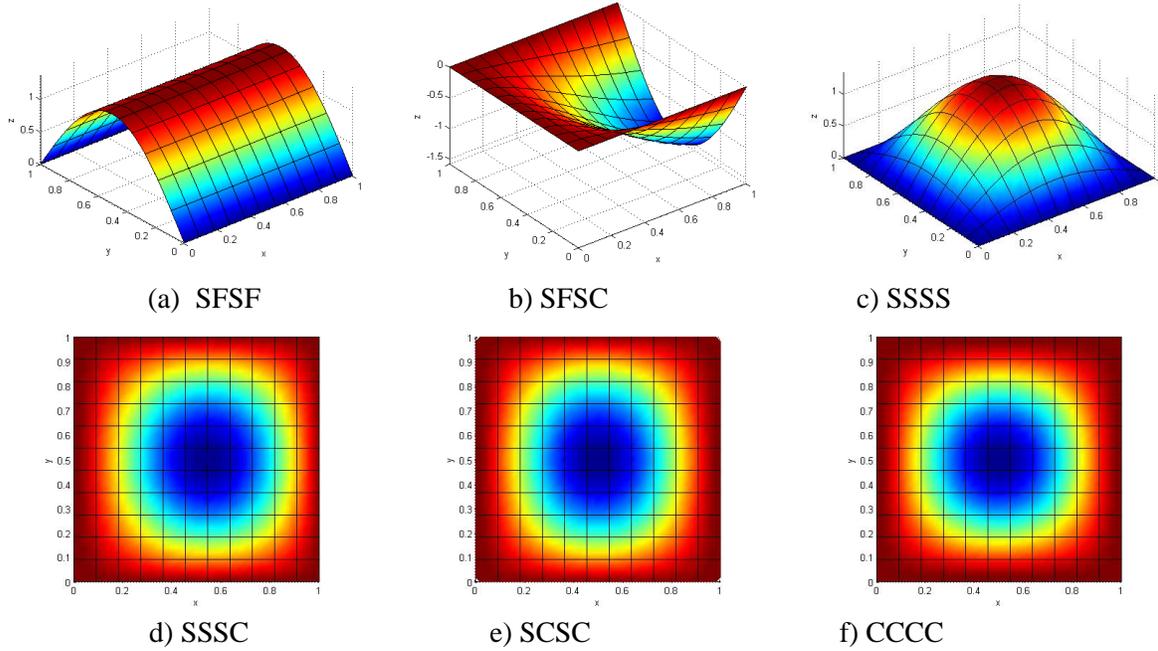

(a) SFSF     b) SFSC     c) SSSS

d) SSSC     e) SCSC     f) CCCC

Figure 6. Mode shape profile of ten layers $[0/90]_5$ composite plate under various boundary conditions

*4.3 Buckling analysis*

A simply supported two-layer angle-ply $[\theta/-\theta]$ square plate is subjected to uniaxial compressive load along *x*-direction shown in Figure 7a. Material set II is used. The results are compared with that of Ren [20] and the analytical solution [26] using FSDT, HSDT and RPT assumptions. For all values of *a/h* ratio and fiber orientation, present model with third-order functions give the closest buckling load to that of RPT predicted by Thai et al [26]. It can be again seen that all models give the same results for thin plates (*a/h*=100). Figure 8 illustrates the buckling mode of two-layer angle-ply composite plate in case of $\theta = 45$. It can be seen that as plate thickness reduces, the non-dimension buckling value $\bar{\lambda}_{cr} = \lambda_{cr} a^2 / E_2 h^3$ increases according to changing of mode shape from two halves sine wave (*a/h*=4) to a half sine wave ($a/h$ =10; 100, respectively). Furthermore, the portion of shear deflection components $w_s$ in transverse displacement reduces and tends to zero as $a/h$=100. The present models, hence, reduce to CLPT model.

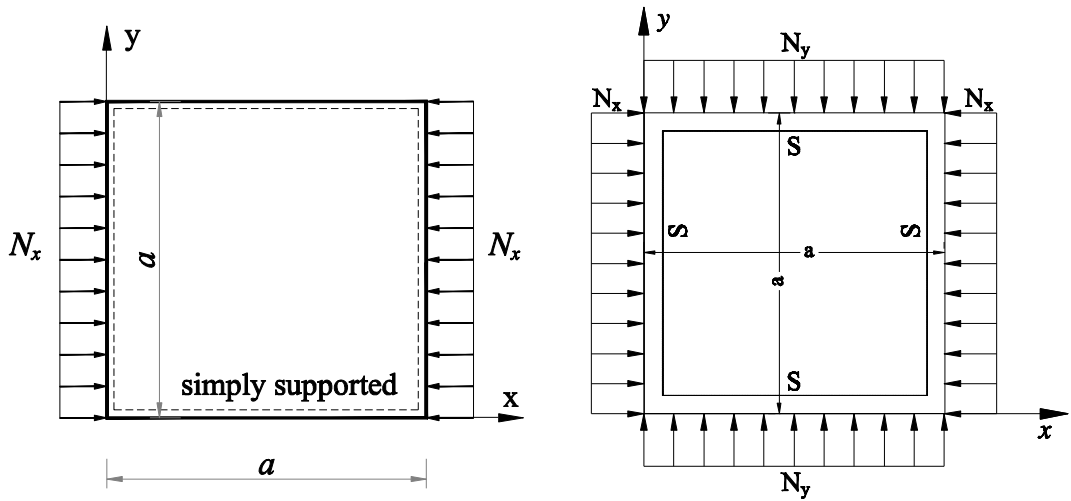

Figure 7. Geometry of laminated composite plates under axial (a) and biaxial (b) compression.

Table 8: The normalized uniaxial buckling load $\bar{\lambda}_{cr}$ of simply supported two layer $[\theta/-\theta]$ composite plate with $E_1/E_2 = 40$.

| | | Ren | Analytical solution [26] | | | RPT-IGA | | | | |
|---|---|---|---|---|---|---|---|---|---|---|
| $a/h$ | $\theta$ | [20] | HSDT | FSDT | RPT | Reddy | Shimpi | Arya | Karama | FiSDT |
| 4 | 30 | 9.5368 | 9.3391 | 7.545 | 9.3518 | 9.3522 | 9.3522 | 9.6731 | 9.9211 | 10.2046 |
| | 45 | 9.82 | 8.2377 | 6.7858 | 8.3963 | 8.3966 | 8.3966 | 8.6472 | 8.9414 | 9.3869 |
| 10 | 30 | 15.7517 | 17.1269 | 16.6132 | 17.2795 | 17.2797 | 17.2797 | 17.3495 | 17.4311 | 17.5489 |
| | 45 | 16.4558 | 18.1544 | 17.5522 | 18.1544 | 18.1545 | 18.1545 | 18.2383 | 18.3354 | 18.4737 |
| 100 | 30 | 20.4793 | 20.5017 | 20.4944 | 20.504 | 20.5042 | 20.5042 | 20.5052 | 20.5063 | 20.5078 |
| | 45 | 21.6384 | 21.6663 | 21.6576 | 21.6663 | 21.6664 | 21.6664 | 21.6676 | 21.6689 | 21.6707 |

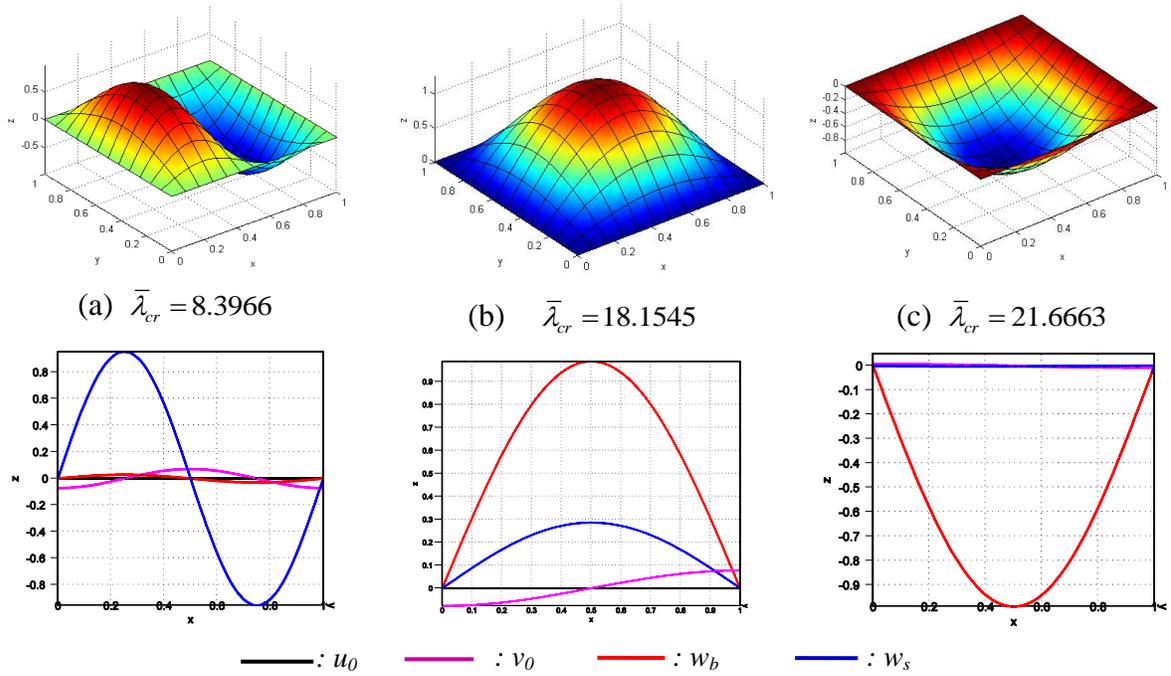

(a) $\bar{\lambda}_{cr} = 8.3966$  (b) $\bar{\lambda}_{cr} = 18.1545$  (c) $\bar{\lambda}_{cr} = 21.6663$

———— : $u_0$   ———— : $v_0$   ———— : $w_b$   ———— : $w_s$

Figure 8. Buckling mode shapes: full plate (upper) and line $y = a/2$ (lower) of simply supported [45/-45] composite plate with $E_1/E_2 = 40$ with various length to thickness ratios: (a) $a/h$=4; (b) $a/h$=10; (c) $a/h$=100.

Table 9: The normalized biaxial buckling load $\bar{\lambda}_{cr}$ of simply supported three-layer [0/90/0] composite plate with $a/h$ =10 and various $E_1/E_2$ ratios.

|  |  | $E_1/E_2$ |  |  |
|---|---|---|---|---|
| Plate model |  | 10 | 20 | 40 |
| FSDT-FEM [52] |  | 4.963 | 7.588 | 10.202 |
| HSDT-FEM [54] |  | 4.963 | 7.516 | 10.259 |
| RPT-IGA | Reddy | 5.1067 | 7.8382 | 10.8825 |
|  | Shimpi | 5.1067 | 7.8382 | 10.8825 |
|  | Arya | 5.1077 | 7.8288 | 10.8549 |
|  | Karama | 5.1105 | 7.8228 | 10.8336 |
|  | FiSDT | 5.1171 | 7.8238 | 10.8247 |

Table 10: The normalized biaxial buckling load $\bar{\lambda}_{cr}$ of simply supported three-layer [0/90/0] composite plate under various $a/h$ =10 ratios.

|  | $a/h$ |  |  |  |
|---|---|---|---|---|
| *Plate model* | 5 | 10 | 15 | 20 |
| HSDT- RPIM [53] | 5.519 | 10.251 | 12.239 | 13.164 |
| FSDT- RPIM [53] | 5.484 | 10.189 | 12.213 | 13.132 |
| HSDT- FEM [54] | 5.526 | 10.259 | 12.226 | 13.185 |
| Reddy | 6.1752 | 10.8825 | 12.714 | 13.5135 |

|   |        |        |         |        |         |
|---|--------|--------|---------|--------|---------|
|        | Shimpi | 6.1752 | 10.8825 | 12.714 | 13.5135 |
| RPT-IGA | Arya  | 6.1571 | 10.8549 | 12.6957 | 13.5014 |
|        | Karama | 6.150  | 10.8336 | 12.6808 | 13.4916 |
|        | FiSDT  | 6.166  | 10.8247 | 12.6728 | 13.4858 |

Finally, we consider a three-layer symmetric cross-ply [0/90/0] simply supported plate subjected to the biaxial buckling load as shown in Figure 7b. Various length-to-thickness $a/h$ and elastic modulus $E_1/E_2$ ratios are studied in this example. Table 9 and Table 10 show the critical buckling parameter $\bar{\lambda}_{cr}$ respect to various modulus and length-to-thickness ratios. The obtained results are compared with those of the finite element formulation based on FSDT [52], the finite element method based on HSDT [54], the mesh free method based on both FSDT and HSDT [53]. The present method shows a good performance compared to other methods for various modulus ratios and length to thickness ratios. The normalized critical biaxial buckling loads are increased with respect to increasing the modulus ratio $E_1/E_2$. Figure 9 reveals the buckling mode shapes of there-layer [0/90/0] composite plate with full plate model and along line $x = a/2$.

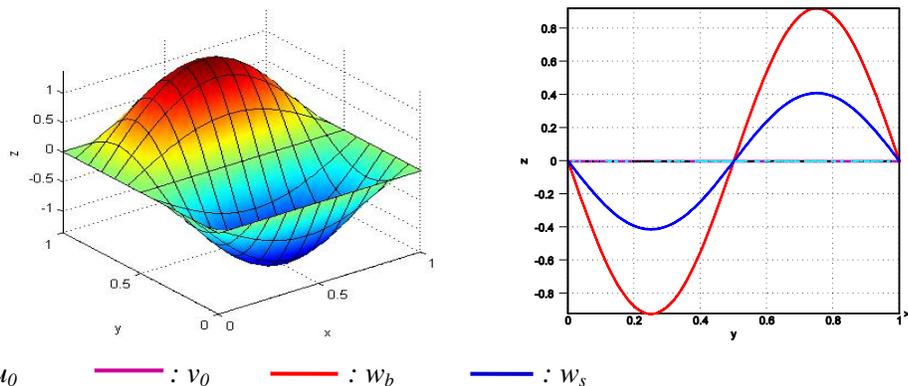

———— : $u_0$    ———— : $v_0$    ———— : $w_b$    ———— : $w_s$

Figure 9. Buckling mode shapes: full plate and line $x = a/2$ of simply supported [0/90/0] composite plate with $E_1/E_2 = 40$, $a/h=10$ (note that $u_0 \equiv v_0$).

## 5. Conclusions

In this paper, the present results by RPT-IGA are compared with the analytical solutions and that of HSDT or 3D elastic models and the excellent agreement in static, free vibration and buckling analysis of laminated composite plates are observed. Using many kinds of higher-order distributed functions, RPT model satisfied naturally the traction-free conditions at the top and bottom plate surfaces. With one variable lower than that of TSDT by Reddy, the present model also ensured the non-linear distribution of the shear stresses/strains through the plate thickness without using SCF. Furthermore, the shear strains/stresses are obtained independently on the bending component. RPT has been thus strongly similar to CLPT. As a consequence, the present results are asymptotic to CLPT ones as plates become thin.

**References**

1. Pagano NJ. Exact solutions for rectangular bidirectional composites and sandwich plates. Journal of Composite Materials 1970;4(1):20–34.
2. Noor AK. Free vibration of multilayered composite plates. AIAA J. 1973;11(7): 1038–1039.
3. Noor AK. Stability of multilayered composite plates. Fiber. Sci. Tech. 1975; 8(2):81-89.
4. Reddy JN. Mechanics of laminated composite plates-theory and analysis. NewYork: CRC Press; 1997.
5. Reissner E. The effect of transverse shear deformation on the bending of elastic plates. J Appl Mech Trans ASME 1945; 12(2):69–77.
6. Mindlin RD. Influence of rotary inertia and shear on flexural motions of isotropic, elastic plates. J Appl Mech Trans ASME 1951; 18(1):31–38.
7. Ferreira AJM, Castro LMS and Bertoluzza S. A high order collocation method for the static and vibration analysis of composite plates using a first-order theory. Compos Struct 2003; 34(7):627-636.
8. Reddy JN. Analysis of functionally graded plates. Int J Numer Methods Eng 2000;47:663–684.
9. Ambartsumian SA. On the theory of bending plates. Izv Otd Tech Nauk ANSSSR 1958;5:269–277.
10. Reissner E. On transverse bending of plates including the effects of transverse shear deformation. IntJ Solids and Structures 1975;25:495–502.
11. Levinson M. An accurate simple theory of statics and dynamics of elastic plates. Mechanics Research Communications 1980;7:343-350.
12. Reddy JN. A simple higher-order theory for laminated composite plates. Journal of Applied Mechanics 1984; 51:745–752.
13. Mantari JL, Oktem AS and Guedes Soares C. Bending response of functionally graded plates by using a new higher order shear deformation theory. Composite Structures 2012;94:714–723.
14. Zenkour AM. The refined sinusoidal theory for FGM plates on elastic foundations. Int J Mech Sci 2009; 51(11–12): 869–880.
15. Ait Atmane H, Tounsi A, Mechab I and Adda Bedia EA. Free vibration analysis of functionally graded plates resting on Winkler-Pasternak elastic foundations using a new shear deformation theory. Int J Mech Mater Des 2010;6(2):113–121.
16. Soldatos KP. A transverse shear deformation theory for homogenous monoclinic plates. Acta Mechanica 1992;94:195–220.

17. Touratier M. An efficient standard plate theory. Int J Eng Sci 1991;29(8):901–916.
18. Karama M, Afaq KS, and Mistou S. Mechanical behavior of laminated composite beam by new multi-layered laminated composite structures model with transverse shear stress continuity. Int J Solids and Structures 2003;40:1525-1546.
19. Arya H, Shimpi RP, and Naik NK. A zigzag model for laminated composite beams. Composite Structures 2002; 56(1):21-24.
20. Aydogdu M. A new shear deformation theory for laminated composite plates. Composite Structures 2009; 89(1):94-101.
21. Senthilnathan NR, Lim SP, Lee KH, and Chow ST. Buckling of Shear-Deformable Plates, AIAA Journal 1987;25(9):1268-1271.
22. Shimpi RP. Refined plate theory and its variants. AIAA Journal 2002; 40(1):137-146.
23. Shimpi RP and Patel HG. A two variable refined plate theory for orthotropic plate analysis, Int J Solids and Structures 2006; 43(22–23); 6783-6799.
24. Shimpi RP and Patel HG. Free vibrations of plate using two variable refined plate theory, J Sound Vib. 2006; 296(4–5): 979-999.
25. Thai HT and Choi DH. A refined plate theory for functionally graded plates resting on elastic foundation. Composites Science and Technology 2011;71(16):1850-1858.
26. Kim SE, Thai HT and Lee J. A two variable refined plate theory for laminated composite plates. Composite Structures 2009;89(2):197-205.
27. Thai HC, Nguyen-Xuan H, Bordas SPA, Nguyen-Thanh N and Rabczuk T. Isogeometric analysis of laminated composite plates using the higher-order shear deformation theory. Mechanics of Advanced Materials and Structures 2012, (in press).
28. Tran V Loc, Ferreira AJM and Nguyen-Xuan H. Isogeometric analysis of functionally graded plates using higher-order shear deformation theory, Composites Part B: Engineering 2013;51: 368-383.
29. Thai H Chien, Tran V Loc, Tran TD, Nguyen-Thoi T and Nguyen-Xuan H. Analysis of laminated composite plates using higher-order shear deformation plate theory and node-based smoothed discrete shear gap method. Applied Mathematical Modelling 2012; 36:5657-5677.
30. Tran V Loc, Nguyen-Thoi T, Thai H Chien, Nguyen-Xuan H. An edge-based smoothed discrete shear gap method (ES-DSG) using the $C^0$-type higher-order shear deformation theory for analysis of laminated composite plates. Mechanics of Advanced Materials and Structures 2012, (in press).
31. Sankara CA and Igengar NGR. A $C^0$ element for free vibration analysis of laminated composite plates. J. Sound and Vib.1996;191:721–738.
32. Kant T and Swaminathan K. Analytical solutions for the static analysis of laminated composite and sandwich plates based on a higher order refined theory. Composite Structures 1996;56(4); 329-344.
33. Hughes TJR, Cottrell JA, and Bazilevs Y. Isogeometric analysis: CAD, finite elements, NURBS, exact geometry and mesh refinement. Comput Methods Appl Mech Eng 2005; 194(39-41):4135–4195.

34. Cottrell JA, Hughes TJR, Bazilevs Y. Isogeometric Analysis, Towards Integration of CAD and FEA. Wiley, 2009.
35. Elguedj T, Bazilevs Y, Calo V, and Hughes T. B and F projection methods for nearly incompressible linear and non-linear elasticity and plasticity using higher-order NURBS elements. Comput Methods Appl Mech Eng 2008;197:2732-2762.
36. Cottrell JA, Reali A, Bazilevs Y and Hughes TJR. Isogeometric analysis of structural vibrations. Comput Methods Appl Mech Eng 2006;195(41-43):5257–5296.
37. Benson DJ, Bazilevs Y, Hsu MC and Hughes TJR. Isogeometric shell analysis: The Reissner–Mindlin shell. Comput Methods Appl Mech Eng 2006;199(5-8): 276–289.
38. Kiendl J and Bletzinger KU, Linhard J and Wchner R. Isogeometric shell analysis with Kirchhoff-Love elements. Comput Methods Appl Mech Eng 2006;198(49-52):3902–3914.
39. Wall WA, Frenzel MA and Cyron C. Isogeometric structural shape optimization. Comput Methods Appl Mech Eng 2008;197(33-40): 2976–2988.
40. Thai HC, Nguyen-Xuan H, Nguyen-Thanh N, Le T-H, Nguyen-Thoi T and Rabczuk T. Static, free vibration, and buckling analysis of laminated composite Reissner–Mindlin plates using NURBS-based isogeometric approach. Int J Numer Meth Engng 2012; 91(6):571–603.
41. Tran V Loc, Thai H Chien, Nguyen-Xuan H, An isogeometric finite element formulation for thermal buckling analysis of functionally graded plates, Finite Element in Analysis and Design 2013;73: 65-76.
42. Nguyen-Thanh N, Kiendl J, Nguyen-Xuan H, Wüchner R, Bletzinger KU, Bazilevs Y and Rabczuk T. Rotation free isogeometric thin shell analysis using PHT-splines. Comput Methods Appl Mech Eng 2011;200(47-48): 3410-3424.
43. Nguyen-Xuan H, Thai HC and Nguyen-Thoi T. Isogeometric finite element analysis of composite sandwich plates using a new higher order shear deformation theory. Composite Part B 2013; 55: 558–574.
44. Auricchio F, Beiraoda Veiga F, Buffa A, Lovadina C, Reali A and Sangalli G. A fully locking-free isogeometric approach for plane linear elasticity problems: a stream function formulation. Comput Methods Appl Mech Eng 2007;197:160–172.
45. Xiao JR, Gilhooley DF, Batra RC, Gillespie JW and McCarthy MA. Analysis of thick composite laminates using a higher-order shear and normal deformable plate theory (HOSNDPT) and a meshless method. Composites Part B: Engineering 2008;39(2):414-427.
46. Vel SS and Batra RC. Analytical solutions for rectangular thick laminated plates subjected to arbitrary boundary conditions. AIAA J 1999;37:1464–73.
47. Khdeir AA and Reddy JN. Analytical solutions of refined plate theories of cross-ply composite laminates. J Pressures Vessel Tech 1991; 113(4):570–8.
48. Whitney JM and Pagano NJ. Shear deformation inheterogeneous anisotropic plates. J Appl Mech, Trans ASME1970;37(4):1031–6.
49. Reddy JN. A simple higher-order theory for laminated composite plates. J Appl Mech,Trans ASME 1984;51:745–52.

50. Kant T and Swaminathan K. Analytical solutions for free vibration of laminated composite and sandwich plates based on a higher-order refined theory. Composite Structures 2001; 53(1):73-85.
51. Reddy JN and Khdeir A. Buckling and vibration of laminated composite plates using various plate theories. AIAA Journal 1989;27(12):1808-1817.
52. Fares ME and Zenkour AZ. Buckling and free vibration of non-homogeneous composite cross-ply laminated plates with various plate theories. Composite Structures 1999;44:279–287.
53. Liu L, Chua LP, and Ghista DN. Mesh-free radial basis function method for static, free vibration and buckling analysis of shear deformable composite laminates. Composite Structures 2007; 78:58–69.
54. Khdeir AA and Librescu L. Analysis of symmetric cross-ply elastic plates using a higher-order theory: PartII: buckling and free vibration. Composite Structures 1988; 9:259–277.

# computers and structures (SCI, IF = 1.5)